# Frenkel-like Wannier-Mott Excitons in Few-Layer PbI$_2$

Alexis S. Toulouse,[1] Benjamin P. Isaacoff,[1] Guangsha Shi,[2] Marie Matuchová,[3] Emmanouil Kioupakis[2] and Roberto Merlin[1]

[1]*Center for Photonics and Multiscale Nanomaterials and Department of Physics, University of Michigan, Ann Arbor, Michigan 48109-1040, USA*

[2]*Department of Materials Science and Engineering, University of Michigan, Ann Arbor, Michigan 48109-2136, USA*

[3]*Institute of Chemical Technology Prague, Technická 1905/5, 160 00 Praha 6-Dejvice, Czech Republic*

Optical measurements and first-principles calculations of the band structure and exciton states in direct-gap bulk and few-layer PbI$_2$ indicate that the $n = 1$ exciton is Frenkel-like in nature in that its energy exhibits a weak dependence on thickness down to atomic-length scales. Results reveal large increases of the gap and exciton binding energy with decreasing number of layers, and a transition of the fundamental gap, which becomes indirect for 1-2 monolayers. Calculated values are in reasonable agreement with a particle-in-a-box model relying on the Wannier-Mott theory of exciton formation. General arguments and existing data suggest that the Frenkel-like character of the lowest exciton is a universal feature of wide-gap layered semiconductors whose effective masses and dielectric constants give bulk Bohr radii that are on the order of the layer spacing.

PACS numbers: 78.67.-n; 78.66.Li; 73.22.-f; 71.35.-y

Following the discovery of graphene [1], two-dimensional systems derived from van der Waals layered materials and, in particular, semiconductors have attracted much attention due to their unusual physical properties and possible applications, including the potential development of a new class of artificial superlattices resulting from the alternate deposition of highly dissimilar substances [2]. Recent work on few-layer semiconductors [3,4,5,6,7,8] has shown that the energy of the lowest exciton associated with the direct gap varies only weakly with layer thickness down to a few layers, a behavior usually associated with highly localized Frenkel excitons. This is in stark contrast with the strong dependence of the lowest direct or indirect gap [3,4,5] and confinement effects observed in semiconductor quantum wells based on, *e. g*., $Al_xGa_{1-x}As$ where the exciton energy and band gaps both increase dramatically with decreasing well width [9]. The observed Frenkel-like behavior is ostensibly in conflict with results indicating that excitons in these materials are not strongly localized; their radii are on the order of a few lattice constants [10,11] and, on that basis alone, one would expect them to fall in a range intermediate between the Frenkel and Mott-Wannier cases. Here we present band-structure calculations, optical reflectance and photoluminescence (PL) measurements on bulk and atomically-thin $PbI_2$, which show that this layered semiconductor follows the pattern observed in other wide-gap layer systems for which both the gap, $E_G$, and the *n* = 1 exciton binding-energy, $E_B$, exhibit large increases with decreasing thickness while the exciton energy, $E_G$-$E_B$, hardly changes. Results are in reasonable agreement with a simple model based on the effective-mass approximation. Relying on this model and available data, we present a plausible explanation as to why the lowest exciton is Frenkel-like and argue that it is a generic property of wide-gap layered semiconductors.

$PbI_2$ is a van der Waals system whose most common polytype, 2*H*, crystallizes in the layered $CdI_2$ structure [12]. As such, it is a good candidate for two-dimensional studies since crystals can be easily cleaved due to the weak inter-layer bonding, and samples of arbitrarily small thickness can be produced [1]. The optical properties and the electronic structure of bulk PbI$_2$ were extensively studied in the late 1960's and early 1970's [11,13,14,15]. Below the fundamental direct gap, $E_0$ = 2.55 eV, experiments at ~ 4 K reveal a prominent quasi-hydrogenic exciton series for which the binding energy of the lowest, $n$ = 1 state is $E_B$ ~ 55 meV [11]. We note that, owing to increased interest in this material for x-ray and γ-ray detection applications, techniques for growing single-crystals of PbI$_2$ have greatly improved in recent years [16].

Few-layer samples were mechanically exfoliated [1] from bulk 2$H$ PbI$_2$ crystals and deposited on silicon wafers covered by a 285-nm-thick oxide layer. Measurements were made over a large range of thicknesses on optically uniform samples. PL measurements were done at 4.5 K. As excitation, we used ~ 15 µW from a 476.5 nm Ar$^+$ laser line. Reflectivity measurements were performed at 77 K using a tungsten-halogen lamp. The incident light was focused using long-working-distance 50× and 100× microscope objectives, which gave spot sizes of ~ 5 and 2 µm in diameter, respectively. Sample thicknesses, given here in units of the $c$ lattice parameter L ≈ 7 Å [12], were estimated from atomic force microscopy. Thickness uncertainties are explicitly indicated for few-layer samples; the estimated error for thicker samples is ~ 5 layers.

Reflectance measurements are shown in Fig. 1 (a). The arrows, labeled FX, indicate the peak energy of the $n$ = 1 exciton gained from simulations using the optical constants of bulk PbI$_2$ [11,13] and the Si/SiO$_2$ substrate, but allowing for small variations of the exciton energy and oscillator strength. The oscillations observed in the samples with 1290 L and 419 L are due to interference effects from multiple reflections. PL spectra are presented in Fig. 1 (b). In the thickest sample we observe three main peaks, two of which, $e$-$h$ (donor-acceptor) and BX (bound exciton), are associated with impurities [17] while the highest-energy band FX is due to free-exciton recombination. The FX energy from PL is in excellent agreement with values from reflectance measurements. Interestingly, the observed emission intensities of *e-h* and BX decrease much more rapidly than FX with decreasing thickness and are not visible in the few-layer crystals.

The most striking feature of the PL and reflectance data is the extremely weak width-dependence of the absolute exciton energy. Aside from small ($\lesssim$ 10 meV) random shifts in the FX position attributed to a sparse presence of the 4*H* polytype [11,17], it is only in samples below ~ 10 monolayers that a significant blue shift ensues as a result of confinement. This shift is clearly visible in the PL spectra of the thinnest samples, which exhibit additional, equally-spaced peaks on the high-energy side we ascribe to forbidden resonant Raman scattering [18] by $A_{2u}$ and $E_u$ longitudinal-optical modes at, respectively, 113 and 106 cm$^{-1}$ [19] and its overtones. The observed larger PL broadening in ultrathin samples is tentatively attributed to enhanced sensitivity to strain from the substrate.

Theoretical band structures of bulk and few-layer PbI$_2$ were obtained from first principles calculations based on density functional theory (DFT) in the local-density approximation [20] using the QUANTUM ESPRESSO code [21] as well as the single-shot GW method [22] using the BerkeleyGW package [23]. Band structures were interpolated with the maximally localized Wannier function method [24]. In addition, exciton eigenstates and eigenvalues were determined using the Bethe-Salpeter-Equation (BSE) code within the BerkeleyGW package [23]. Spin-orbit coupling (SO) effects on the band structures were considered in a non-self-consistent way, as in Ref. [25], but they were ignored in the evaluation of the exciton wavefunctions. Due to the inherent difficulty in accounting for the van der Waals interaction, which determines the interlayer separation, we used atomic-position data for bulk crystals [12]. Parameters such as, *e. g*., the plane-wave cutoff energies and the *k*-grid sampling were chosen to converge the band gap and exciton energies separately to within ~ 0.1 eV.

*Ab initio* band structures are shown in Fig. 2. For simplicity, we plot only the 6 highest valence and 3 lowest conduction bands per layer, which derive predominantly from iodine 5*p* and lead 6*p* orbitals, respectively. Throughout most of the Brillouin zone, these states are separated by a few eV from other bands. Direct band gaps and corresponding *k*-space sampling used are listed in Table I. For bulk PbI$_2$, we find a fundamental direct gap of 2.38 eV at the A point, which agrees relatively well with the experimental value, $E_0 = 2.55$ eV, and previous calculations with empirical pseudo-potentials [14]. The calculated electron (hole) effective masses at the A-point, perpendicular and parallel to the *c*-axis are $m_e^\perp = 0.21$ ($m_h^\perp = 0.59$) and $m_e^\parallel = 1.05$ ($m_h^\parallel = 0.56$), in units of the electron mass; the relevant states involve primarily $p_z$-orbitals. We briefly emphasize the importance of SO effects, which not only shrink the gap by $\approx 0.80$ eV, as in Fig. 2, but also lead to a mixing of states that transforms the character of the direct-gap transition from dipole-forbidden (without SO coupling) to optically allowed.

Large increases in the gap due to quantum confinement are clearly observed in Fig. 2; note that the bulk A-point projects onto the Γ-point of the two-dimensional zone. The most significant changes occur for iodine-like states, reflecting the strong effect neighboring layers have on these atoms because of their position in the layers. Confinement effects are most prominent for the top $p_z$-like valence band, which develops a minimum at the Γ-point for a single monolayer such that the fundamental gap becomes indirect. The bilayer structure also results in an indirect gap slightly smaller than its direct gap; however, the difference is below the accuracy of the calculation. Interestingly, this behavior is the reverse of that of MoS$_2$ for which the gap is indirect except for the monolayer [3].

Central to our arguments concerning the Frenkel-like behavior of the lowest exciton are the properties of its wave function. Results of BSE calculations are shown in Fig. 3; as noted, they do not include the SO interaction although we believe that its inclusion would not significantly modify the exciton radii. Fig. 3 (a) shows the modulus squared of the $n = 1$ bulk wave function Ψ for a fixed hole position [26]. The calculated in- and out-of-plane radii of the ellipsoidal envelope are 18 Å and 12 Å, which are very close to experimental values obtained assuming an isotropic mass tensor [13,27], and a factor of, respectively, 4 and 1.7 larger than the corresponding lattice constants [28]. As expected for hydrogenic systems, confinement enhances the Coulomb interaction as well as the binding energy, reducing in turn the exciton radii, and thereby causing a transition from borderline Wannier-Mott to Frenkel type; see Fig. 3 (b). It is interesting to note that the calculated size reduction from bulk to monolayer $PbI_2$ is a factor of approximately four, as for the ratio between three- and two-dimensional hydrogen [29]. We also observe that, while the $c$-axis bulk radius is slightly less than the thickness of two atomic layers, the calculations indicate that the exciton wave function involves states with wavevectors that are spread over a width of ≈ 10% of the size of the Brillouin zone from the A-point and, thus, well within the range where the bands can be treated as parabolic.

The thickness-dependence of the energy of the $n = 1$ exciton from PL experiments and that of the direct gap from GW+SO results are plotted in Fig. 4. The plot includes also SO-corrected values for the exciton from BSE calculations, obtained by subtracting the SO-induced red shift of the gap. This procedure is justified on the grounds that the introduction of SO coupling leads to a fairly rigid shift of the relevant bands by ~ 0.8 eV (see Fig. 2 and Table I). Moreover, to correct for what we believe is a systematic layer-thickness-independent error due to the approximations involved in our first-principles calculations, all the theoretical results have been rigidly shifted upwards by 0.17 eV [30]. This value represents the difference between the measured [11] and calculated bandgap for bulk PbI$_2$ (note that, unlike the absolute energy, the calculated bulk exciton binding energy of 70 meV is in reasonable agreement with $E_B = 55$ meV from experiments [11]). It is apparent that, on account of this shift, the calculated exciton energies are in excellent agreement with the PL data.

Also shown in Fig. 4 are results of effective-mass calculations for a single electron-hole pair in an infinite square well for which the gap is $E_G = E_0 + \hbar^2\pi^2/2\mu^{\parallel}d^2$, where $d$ is the sample thickness, $\mu^{\parallel} = m_e^{\parallel}m_h^{\parallel}/(m_e^{\parallel}+m_h^{\parallel})$ is the reduced mass and, as before, $E_0$ is the bulk gap. The electron-hole Coulomb interaction is $e^2/\varepsilon r$, where $r$ is the relative coordinate and $\varepsilon$ is the screening constant ($\varepsilon_{\parallel} = 6.25$ and $\varepsilon_{\perp} = 26.75$ in PbI$_2$ [19]). The exciton binding energy was calculated using a variational approach valid for GaAs-Al$_x$Ga$_{1-x}$As quantum wells [31]. This procedure gives values of the binding energy, which are in agreement with the exact results $2e^2/\varepsilon a_0$ for the two-dimensional limit [29] and $e^2/2\varepsilon a_0$ for $d \gg a_0$; $a_0$ is the bulk exciton radius. The results for $\varepsilon = \infty$ (non-interacting pair) agree relatively well with GW+SO calculations for the gap down to two monolayers whereas those for $\varepsilon = 9.5$ are in very good agreement with experimental and theoretical exciton energies down to ~ 3 layers. Moreover, the model predicts negligible (i.e., less than 10 meV) confinement effects for samples thicker than ~ 15 layers. As the layer thickness approaches the atomic limit, we expect effective-mass models to break down and, moreover, exciton parameters to become more and more affected by the surrounding dielectric [4,6]. This applies in particular to mono- and bilayer crystals for which the square-well model strongly underestimates the binding energy.

The totality of our results suggests a simple explanation as to why the lowest exciton in PbI$_2$ and other wide-gap layer systems behaves in a Frenkel manner with decreasing thickness. First, we emphasize the fact that, while the approximation is strictly valid in the Wannier-Mott limit, effective-mass calculations are in good agreement with the PL data and GW+SO results down to atomically thin crystals. It is easy to see that, according to the exciton-in-a-box model, the exciton energy in the limits $d \ll a_0$ and $d \gg a_0$ are, respectively, $E_0 - 4E_B + \hbar^2\pi^2/2d^2\mu^{\parallel}$ [31] and $E_0 - E_B + \hbar^2\pi^2/2d^2M^{\parallel}$ [32]. Here, $M^{\parallel} = m_e^{\parallel} + m_h^{\parallel}$ is the $c$–axis exciton mass. From these expressions, we obtain the crossover length $\ell \sim a_0(1 - \mu^{\parallel}/M^{\parallel})^{1/2} \sim a_0$ and, thus, the thickness below which confinement effects become important. It follows that the observed Frenkel-like behavior is simply the result that the interlayer thickness is on the order of the $c$-axis bulk Bohr radius.

A. S. T. acknowledges support from DOD through the National Defense Science & Engineering Graduate Fellowship Program. G. S. and E. K. were supported from the Center for Solar and Thermal Energy Conversion, an EFRC funded by DOE under Award DE-SC0000957 (band calculations) and from NSF Career Award DMR-1254314 (exciton calculations). This work used resources of the National Energy Research Scientific Computing Center, supported by DOE under Contract No. DE-AC02-05CH11231.

# FIGURE CAPTIONS

**FIG. 1** (color online). (a) Reflectance spectra at 77 K. FX denotes the position of the $n = 1$ exciton. Fabry-Perot oscillations are seen in the two thickest samples. Red-dashed curves are results of simulations; see text. (b) PL data at 4.5 K. Spectra show free (FX) and bound (BX) exciton recombination, and emission due to donor-acceptor pairs (e-$h$).

**FIG. 2** (color online). Band structures without (GW; dashed red) and with (GW+SO; solid blue) spin-orbit coupling. Results are shown along principal directions in the three- (two-) dimensional hexagonal Brillouin zone of bulk (few-layer) $PbI_2$, shown in the inset. The fundamental gap is direct (indirect) for bulk and 3L- (1L- and 2L-) $PbI_2$; calculated values are listed in Table I.

**FIG. 3** (color online). (a) Probability distributions for directions parallel ($a$-axis) and perpendicular ($c$-axis) to the layers for bulk $PbI_2$, with standard deviations (exciton radii) $\sigma_a$ and $\sigma_c$. The central panel shows an isosurface corresponding to the value of $|\Psi|$ at the exciton radii. Black and purple balls represent, respectively, lead and iodine ions. The dashed-red curve is an ellipse centered at the position of the hole whose principal axes are the radii. (b) Radii vs. number of layers. Bulk values are indicated by arrows.

**FIG. 4** (color online). Calculated results for the direct gap (red circles) and the $n = 1$ exciton (blue squares) of few-layer $PbI_2$. The theoretical data have been rigidly shifted upwards by 0.17 eV to match the band gap of bulk $PbI_2$; see text for an explanation. Arrows give bulk theoretical values. Orange diamonds are free-exciton data from PL experiments. Also shown are effective-mass-approximation predictions for the lowest-energy exciton and unbound electron-hole pairs ($\varepsilon = \infty$) in an infinite square well.

**TABLE I.** Direct gaps (eV) from DFT, GW, and GW with spin-orbit coupling (GW+SO) calculations and the *k*-grid used to sample the Brillouin zone. Gaps in bulk and few-layer PbI$_2$ are, respectively, at the A and Γ point.

| Structure | DFT | GW | GW+SO | *k*-grid |
|:---:|:---:|:---:|:---:|:---:|
| bulk | 1.99 | 3.18 | 2.38 | 6×6×4 |
| 4 layer | 2.07 | 3.49 | 2.67 | 10×10×1 |
| 3 layer | 2.12 | 3.60 | 2.78 | 9×9×1 |
| 2 layer | 2.24 | 3.82 | 3.01 | 8×8×1 |
| 1 layer | 2.70 | 4.56 | 3.72 | 8×8×1 |

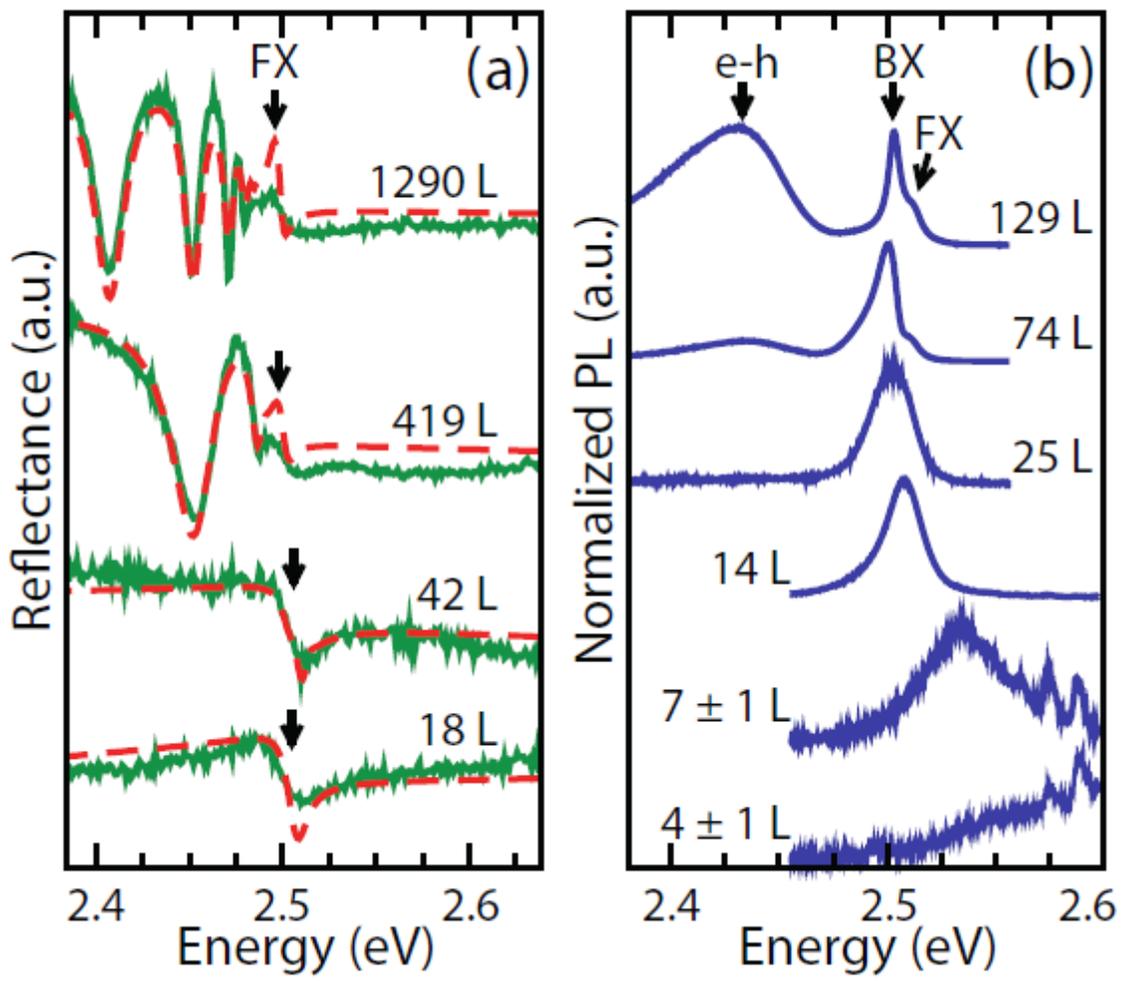

Figure 1

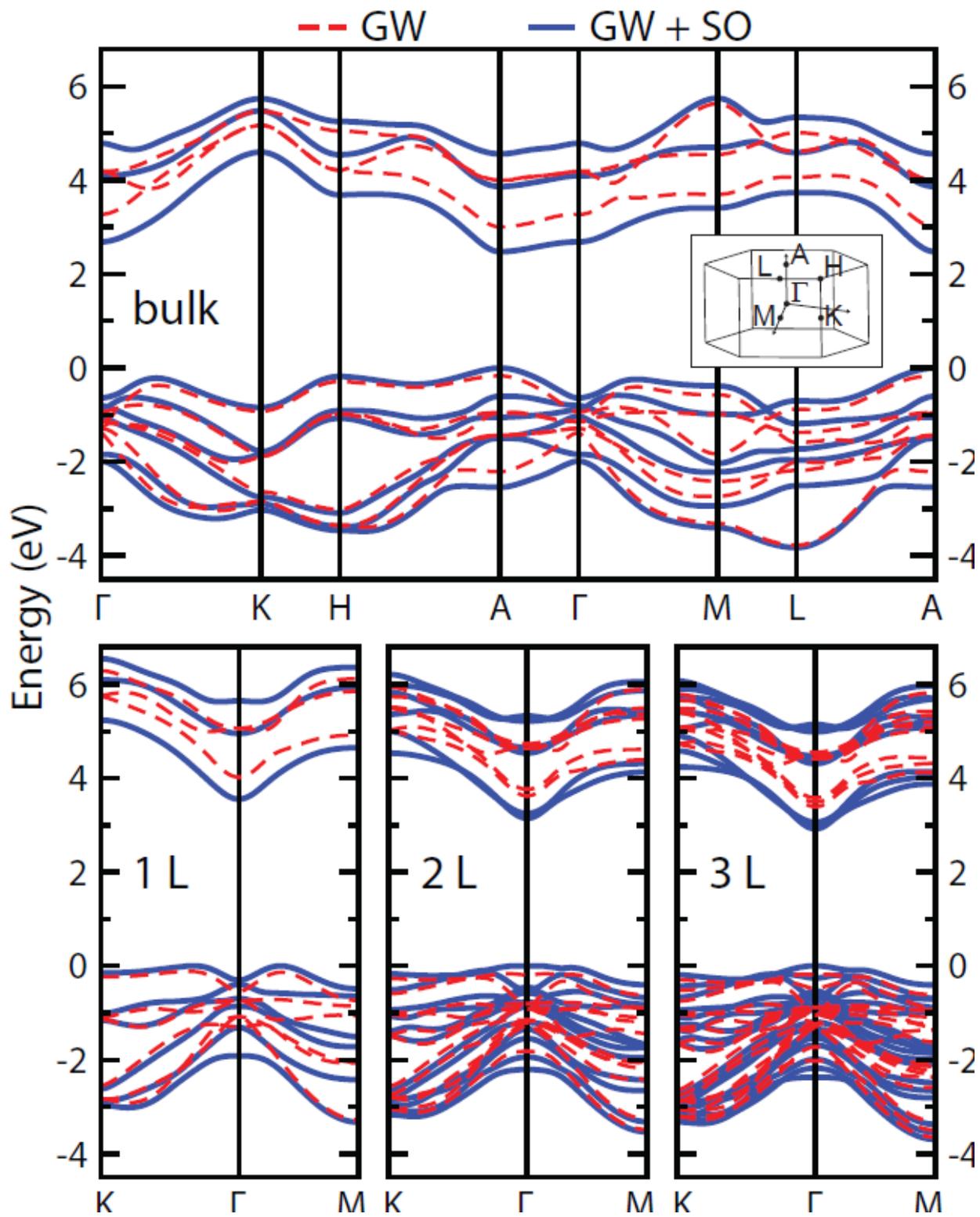

Figure 2

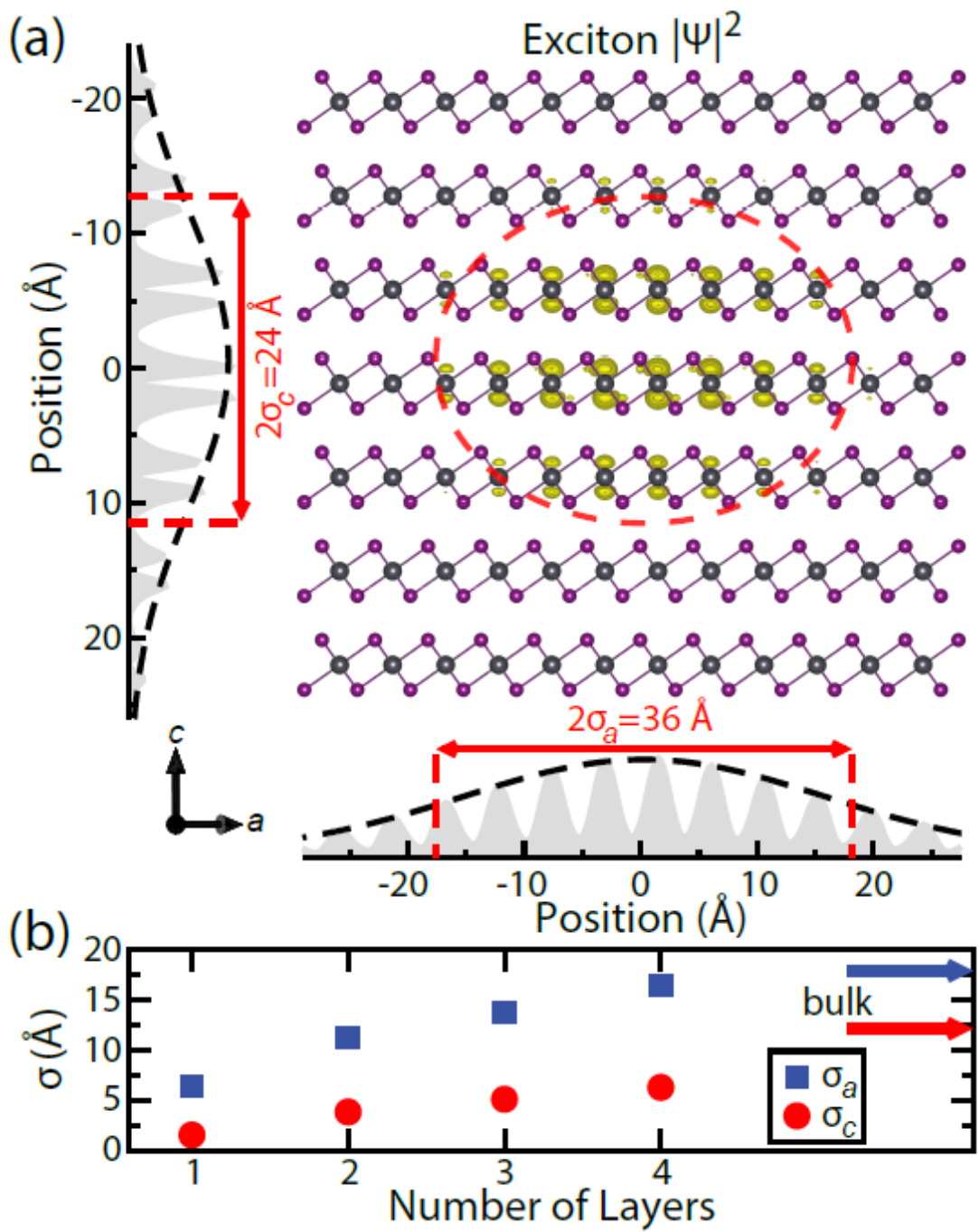

Figure 3

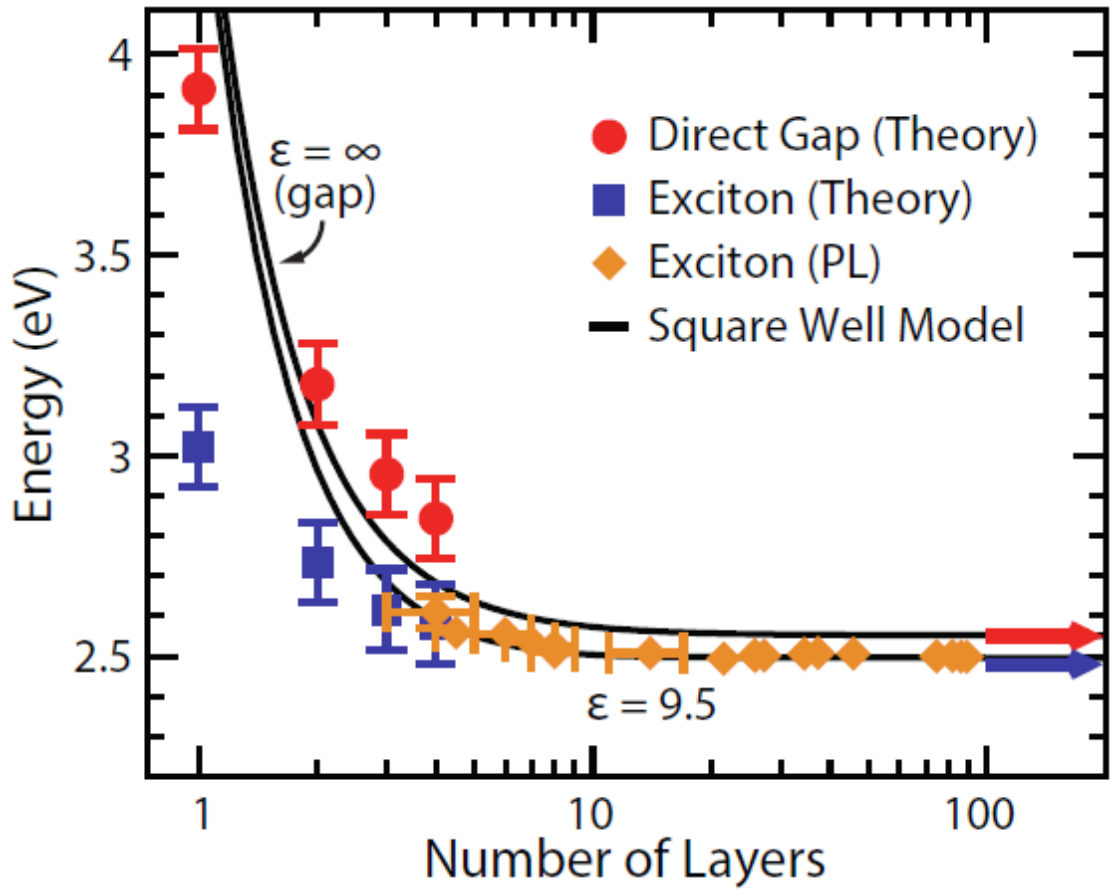

Figure 4